\documentclass[amstex,onecolumn,showpacs,floats,floatfix,aps,pra]{revtex4}
\usepackage{amssymb}
\usepackage{amsmath}
\usepackage{calc}
\usepackage{graphicx}
\usepackage{bm}

\begin{document}

\author{Mouhamad Al-Mahmoud}
\affiliation{Department of Physics, Sofia University, James Bourchier 5
blvd, 1164 Sofia, Bulgaria}
\author{Andon Rangelov}
\affiliation{Department of Physics, Sofia University, James Bourchier 5
blvd, 1164 Sofia, Bulgaria}
\author{Virginie Coda}
\affiliation{Universit\'e de Lorraine, CentraleSup\'elec, LMOPS, F-57000
Metz, France}
\author{Germano Montemezzani}
\affiliation{Universit\'e de Lorraine, CentraleSup\'elec, LMOPS, F-57000
Metz, France}

\title{Segmented Composite Optical Parametric Amplification}
\date{\today }

\begin{abstract}
We propose a novel optical parametric amplification scheme which combines quasi-phase-matching with a composite
pulse approach that involves crystal segments of specific lengths. The presented scheme highly increases the robustness of the frequency conversion against variations of the nonlinear coupling and of the pump, idler or signal wavelengths and has therefore the potential to enhance high amplification and broadband operation. Simulations examples applied to LiNbO$_3$ are given.
\end{abstract}

\maketitle

\section{Introduction}
Some optical applications require short optical pulses with large peak power, which may be obtained with the help of optical parametric amplifiers (OPAs) \cite{Joosen,Gale1995,Gale1998,Schmidt} that are among the most useful nonlinear optical devices. Optical parametric amplification consists in the nonlinear interaction of three waves. In this process the two waves at the longer wavelengths, the input signal wave as well as the idler wave, gain power at the expense of the pump wave being at the shortest wavelength. In OPA the main obstacle encountered when short pulses
are used is to combine a high signal amplification and a sufficiently broad amplification bandwidth. The latter is limited because material dispersion imposes that, for a given wave interaction configuration, the exact phase-matching condition can be strictly satisfied only for a single set of wavelengths of the three waves.
Presently the most common way to achieve the combination of broadband and high amplification is to use chirped quasi-phase-matching
\cite{Harris,Charbonneau-Lefort1,Charbonneau-Lefort2,Phillips2012,Phillips2013,Moses}.
Even though chirped quasi-phase-matching approaches have the advantage of
being broadband, they require high pump input intensity and/or very long
nonlinear crystals. Very high pump intensities have the drawback of a possible damage of the nonlinear crystal when they approach its
damage threshold.

In this paper we explore an alternative method to achieve broadband
amplification bandwidth together with high amplification. The technique involves a combination of quasi-phase-matching (QPM) gratings together with a segmentation of the crystal that implements the equivalent to the composite pulses approach used in Nuclear Magnetic Resonance (NMR) to prepare given quantum states in a robust way
\cite{Levitt,Shaka1985,Shaka1987}. The present approach leads to a highly increased robustness of the nonlinear amplification process with respect to both, the phase mismatch (associated to a change of wavelength or of temperature) and the coupling strength.
Section 2 gives the general theory while Sect. 3 describes the numerical approach and the simulation results obtained with the best crystal segmentation. Some practical examples are given for the important case of MgO-doped quasi phase-matched LiNbO$_3$. Finally, Sect. 4 summarizes.

\section{Theory}

We start with the simmetrized coupled wave equations for collinear
three-wave mixing in the slowly varying envelope approximation \cite%
{Boyd,Yariv}
\begin{subequations}
\label{three-wave mixing}
\begin{align}
i\partial _{z}A_{1}& =\widetilde{\Omega }A_{2}^{\ast }A_{3}\exp \left[
-i\Delta kz\right] , \\
i\partial _{z}A_{2}& =\widetilde{\Omega }A_{1}^{\ast }A_{3}\exp \left[
-i\Delta kz\right] , \\
i\partial _{z}A_{3}& =\widetilde{\Omega }A_{1}A_{2}\exp \left[ i\Delta kz%
\right] ,
\end{align}
\end{subequations}
where $\widetilde{\Omega }=-(2\chi ^{(2)}/ \pi c )\sqrt{\omega _{1}\omega
_{2}\omega _{3}/n_{1}n_{2}n_{3}}$ is the effective nonlinear coupling coefficient for first-order QPM, $z$ is the
position along the propagation axis, $\omega _{j}$ are the frequencies of the three involved waves and  $n_{j}$ are their refractive indices.  Here $j=1,2,3$ refer to the
signal, idler and pump waves, respectively.  The quantity  $\chi ^{\left( 2\right) }$ in $\widetilde{\Omega}$
is the effective second-order susceptibility and $c$ is the speed of light in vacuum.  The amplitudes $A_{j}\equiv
\sqrt{n_{j}/\omega _{j}}\,E_{j}$ in (\ref{three-wave mixing}) are proportional to the amplitudes $E_{j}$
of the wave electric fields; $|A_{j}|^{2}$ is proportional to the number of
photons associated to the $j$th wave. It is important to note that Eq. (\ref{three-wave mixing}) is written in a form that assumes that quasi-phase-matching is implemented and that the quasi-phase matching period is sufficiently short as compared to the interaction length. Therefore the phase mismatch parameter $\Delta k$ already contains the mismatch compensation term associated to the periodic grating, that is
\begin{equation}
\Delta k=k_{1}+k_{2}-k_{3}+2\pi/\Lambda \equiv \widetilde{\Delta k} + 2\pi/\Lambda   .
\label{Delta-k}
\end{equation}
where $\Lambda$ is the quasi-phase-matching period, that is the first order local poling period in the case of periodically poled crystals. Obviously, for the central operation wavelengths at which the device is designed one has $\Delta k=0$. At the same time the true phase mismatch $\widetilde{\Delta k}=k_{1}+k_{2}-k_{3}$, which depends only on the wave-vectors $k_j$ of the three interacting waves, is generally quite far from vanishing. The set of equations (\ref{three-wave mixing}) could have been written also by using the quantities $\widetilde{\Delta k}$ instead of $\Delta k$, however in this case $\widetilde{\Omega}$ would need to switch its sign after each distance $\Lambda/2$ and the term $2/\pi$ would need to be dropped in the nonlinear coupling coefficient.

\begin{figure}[!htb]
\centering
\includegraphics[width=0.5\columnwidth]{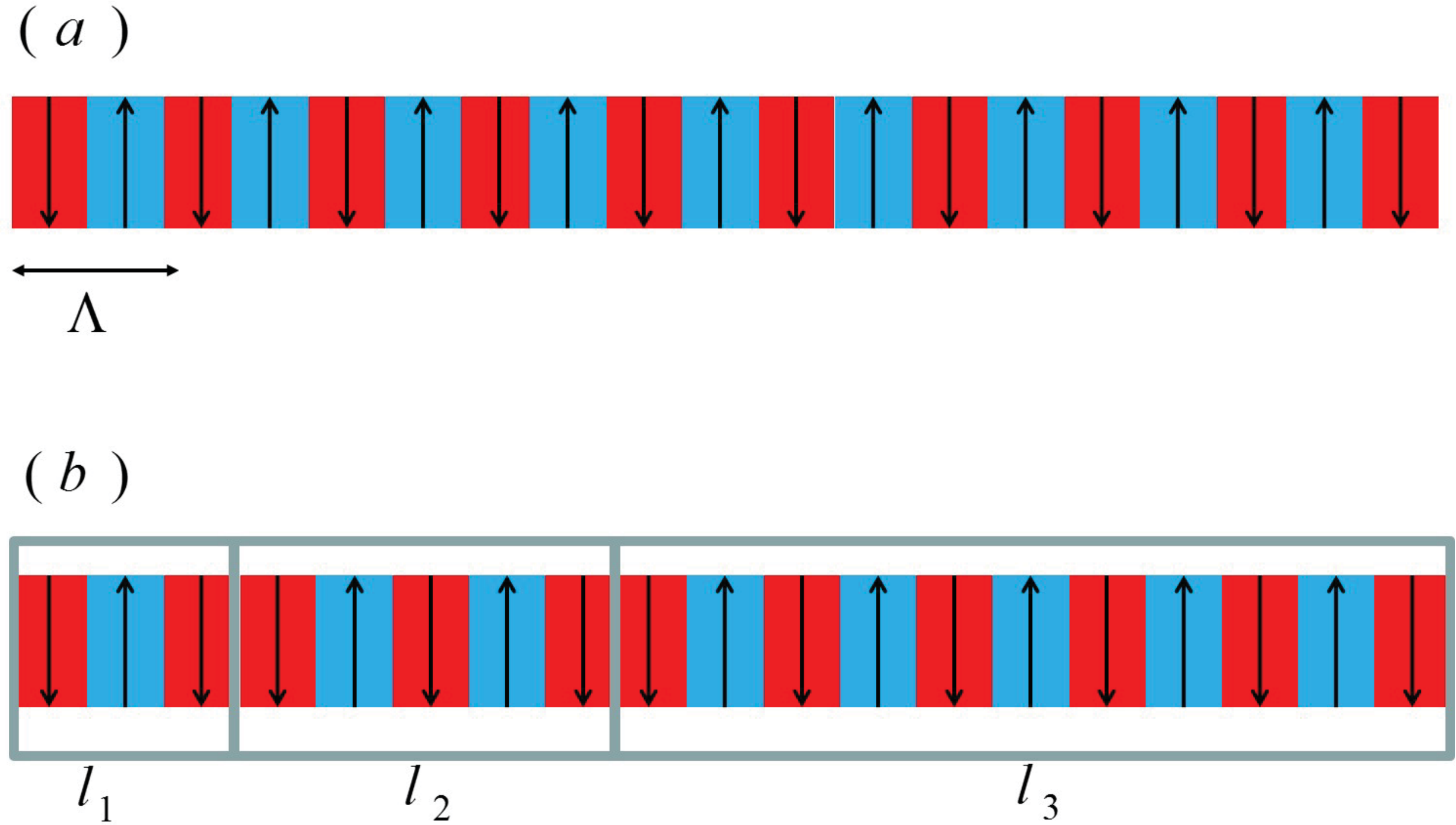}
\caption{Figure depicting sign reversal of $\protect\chi ^{(2)}$ nonlinear
coefficient for (a) Standard quasi-phase matching (QPM) technique with local
modulation period $\Lambda $. (b) Composite segmented periodically
poled design, with example of 3 segments. All segments are periodically
poled with the same period $\Lambda $, however once a new segment begins,
the sign reversal order of $\protect\chi ^{(2)}$ is changed.}
\label{fig1}
\end{figure}

Depending upon the initial conditions $%
A_{j}\left( z=0\right) $, different processes can arise: sum frequency generation
(SFG), difference frequency generation (DFG) or OPA. Here we consider the OPA case, we assume that $\omega _{3}=\omega
	_{1}+\omega _{2}$  and we treat first Eq. (\ref{three-wave mixing}) in the limit of validity of the undepleted pump approximation ($A_{3}=$%
~const). In this limit the set of equations (\ref{three-wave mixing}) turns into
\begin{subequations}
\label{OPA}
\begin{align}
i\partial _{z}A_{1}& =\Omega A_{2}^{\ast }\exp \left[ -i\Delta kz\right] , \\
i\partial _{z}A_{2}& =\Omega A_{1}^{\ast }\exp \left[ -i\Delta kz\right] ,
\end{align}
\end{subequations}
with $\Omega =\widetilde{\Omega }A_{3}$ being a modified coupling coefficient. When the wave vector mismatch $%
\Delta k$ and the coupling coefficient $\Omega $ are constant then equations (%
\ref{OPA}) possess exact analytic solutions \cite{Shen}
\begin{subequations}
\begin{align}
A_{1}\left( z\right) & =e^{i\Delta kz}\left[ A_{1}\left( 0\right) \left(
\cosh \left( gz\right) -\frac{i\Delta k}{2g}\sinh \left( gz\right) \right) +%
\frac{i\Omega }{g}A_{2}^{\ast }\left( 0\right) \sinh \left( gz\right) \right]
, \\
A_{2}\left( z\right) & =e^{i\Delta kz}\left[ A_{2}\left( 0\right) \left(
\cosh \left( gz\right) -\frac{i\Delta k}{2g}\sinh \left( gz\right) \right) +%
\frac{i\Omega }{g}A_{1}^{\ast }\left( 0\right) \sinh \left( gz\right) \right]
,
\end{align}
\end{subequations}
where $g=\sqrt{\Omega ^{2}-\left( \frac{\Delta k}{2}\right) ^{2}}$ is a gain coefficient. From the
last equations one can easily see that both signal and idler increase
exponentially for large $\Omega z$ and in the case $\Delta k=0$%
\begin{subequations}
\label{solutions}
\begin{align}
\left\vert A_{1}\left( z\right) \right\vert & \approx \frac{\left\vert
A_{1}\left( 0\right) \right\vert }{2}\exp \left[ \Omega z\right] , \\
\left\vert A_{2}\left( z\right) \right\vert & \approx \frac{\left\vert
A_{2}\left( 0\right) \right\vert }{2}\exp \left[ \Omega z\right] .
\end{align}
\end{subequations}
In this work we are interested in an optimization of the signal intensity amplification factor $a$, defined as the ratio of the intensity of the signal wave ($A_1$) taken at distance $z$ to its intensity at the entrance of the crystal
\begin{equation}
a=\frac{\left\vert A_{1}\left( z\right) \right\vert ^{2}}{\left\vert
A_{1}\left( 0\right) \right\vert ^{2}}=\frac{I_{1}\left( z\right) }{%
I_{1}\left( 0\right) }.  \label{parametric gain}
\end{equation}

The above argumentation in connection with Eq. (\ref{solutions}) indicates that, when the phase-matching condition is
satisfied ($\Delta k =0$), the OPA process is the most efficient. However this is not
entirely true, because the solutions (\ref{solutions}) are derived only in the
limit of the undepleted pump approximation. If one considers the depleted pump regime, the nonlinear equations (\ref{three-wave mixing}) have
solutions in terms of Jacobi elliptic functions \cite{Baumgartner,Wolfram}.
These, like trigonometric functions, are periodic
functions and thus the energy transfer oscillates back and forth between
pump field and signal field. The chirped quasi-phase-matching techniques \cite%
{Harris,Charbonneau-Lefort1,Charbonneau-Lefort2,Phillips2012,Phillips2013}, in addition to the improvement of the bandwidth,
eliminate the problem of back conversion and can be used even in the case of the depleted pump regime.

Here, in analogy with the technique of composite pulses from quantum physics
\cite{Levitt,Shaka1985,Shaka1987,Levitt2,Freeman,Blatt, Wunderlich, Torosov,
Schraft}, we propose to use segmented composite crystals for OPA. We note that the
composite pulse analogy was already used in nonlinear optics but for the sum frequency generation or second harmonic generation (SHG) \cite%
{Genov,Rangelov,Erlich}. In these cases, in the undepleted pump approximation the
differential equations governing the spatial dynamics have a SU(2) symmetry \cite%
{Suchowski1,Suchowski2}, which is exactly the same symmetry possessed by quantum
systems with two states \cite{Torosov, Schraft}. Therefore the mapping
between two-state quantum systems and SFG in the undepleted pump regime is complete and one
can use the known analytic solutions from quantum physics to find robust
solutions in nonlinear optics \cite{Genov,Rangelov,Erlich}. In the case of OPA there is no SU(2) symmetry and it is thus not possible to exploit known composite pulses analytic solutions. Instead, we are going to derive numerically solutions that
achieve broadband amplification bandwidth together with high amplification in the depleted pump regime.

%***************************************************************
\begin{table}[t]
\centering%
\begin{tabular}{|c|c|c|}
\hline
$N$ &Name& Segment lengths $l_{1};l_{2};\ldots ;l_{N}$ in units of $L$ \\ \hline
3 & 3 & 0.373; 0.594; 0.033 \\ \hline
4 & 4 & 0.303; 0.522; 0.124; 0.051 \\ \hline
6 & 6a & 0.293; 0.258; 0.003; 0.255; 0.124; 0.067 \\ \hline
6 & 6b & 0.168; 0.035; 0.345; 0.023; 0.222; 0.207 \\ \hline
6 & 6c & 0.223; 0.005; 0.404; 0.175; 0.113; 0.080 \\ \hline
8 & 8 & 0.022; 0.064; 0.046; 0.205; 0.270; 0.096; 0.222; 0.075 \\ \hline
\end{tabular}%
\caption{Numerically found segment lengths $l_i$ (in units of total crystal length
$L$) for composite segmented periodically poled design with $N$ segments. The given values for $l_i$
are such as to optimize the robustness of the OPA process against variations of the nonlinear coupling coefficient and of the phase mismatch $\Delta k$.}
\label{Table}
\end{table}
%***************************************************************

%%%%%%%%%%%%%%%%%%%%%%%%%%%%%%%%%%%%%%%%%%
\section{General numerical approach and LiNbO$_{3}$ crystal simulations}

The procedure that we track is the following: the period of the flip sign of
the nonlinear susceptibility $\chi ^{\left( 2\right) }$ is such that the phase mismatch for the OPA process will
be zero in Eq. (\ref{Delta-k}) due to QPM, resulting in a local modulation period $%
\Lambda$ (Fig.\ref{fig1} (a)). However, in contrast to the periodic design, we induce additional
sign flips of the coupling coefficient at specific boundaries. In practice at each
segment boundary there are two domains with the same orientation that merge
in a single double-as-long domain, as shown in Fig.\ref{fig1} (b). The
periodic sign switch of $\chi ^{\left( 2\right) }$ ensures the phase matching for
OPA ($\Delta k=0$) and the additional sign switches of $\chi ^{\left( 2\right) }$ at the segment boundaries will
change the sign of $\widetilde{\Omega }$ in the whole crystal segment in a similar fashion as in
Shaka-Pines pulses from NMR \cite{Shaka1985,Shaka1987}. We denote the intervals between two double
length domains as $l_{1},l_{2},l_{3}...l_{N}$ as shown in Fig.\ref{fig1} (b)
and we find each length $l_{k}$ using Monte Carlo simulations. We set
our units of length to be the crystal length $L$, then we allow the coupling
$\widetilde{\Omega }$ to vary from zero to $30/L$ and the phase mismatch $\Delta
k$ to vary from $-15/L$ to $15/L$. In this way we generate 10$^{5}$ sets of random
segments $l_{1},l_{2},l_{3}...l_{N}$. Finally we pick up solutions, which 2D integral of the amplification $a$ in the above interval of $\Delta k$ and $\widetilde{\Omega }$ deliver the highest value. The lengths of composite segments $%
l_{1},l_{2},l_{3}...l_{N}$ for the best solutions are listed in Table \ref%
{Table}. We have found that the use of a small number of composite segments (two, three
and partly four) do not lead to any strong improvement with respect to the standard QPM case.
In contrast, already for a moderate number of segments between six and eight we find a significant improvement of the robustness of the amplification process. In this case our analysis shows that there are different solutions for the optimum segment configurations (given in Table \ref%
{Table}), which works better depending on the initial amplitude of the signal wave $A_{1}(0)$ (as compared to the pump wave amplitude taken as $A_{3}(0)=1$).

%***************************************************************
\begin{figure}[!htb]
\centering
\includegraphics[width= 0.7\columnwidth]{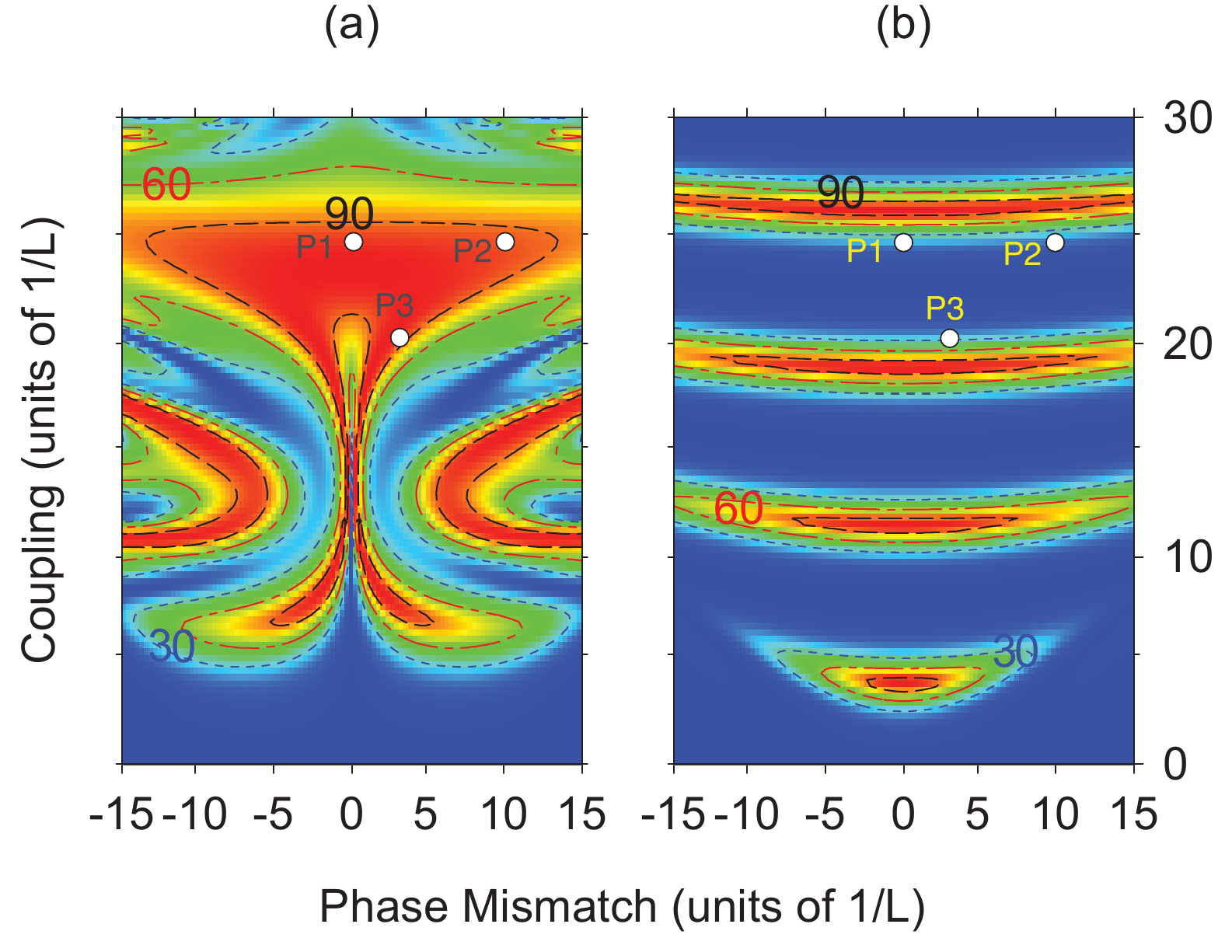}
\caption{(Color online) Signal intensity amplification vs. the phase mismatch $\Delta k$ and the
coupling $\widetilde{\Omega}$ . (a) Composite crystal with six segments (6a from Table
\protect\ref{Table}). (b) Standard periodic QPM design. In both plots the red color indicate area with high signal intensity amplification, while blue color indicate area with low signal intensity amplification. The points P1, P2 and P3 are selected positions used to illustrate the behavior in Fig. \ref{Fig2bis}. The  three isolines mark intensity amplification levels of 30, 60 and 90. The input wave amplitudes are $A_{1}\left( 0\right) =0.1$, $A_{2}\left( 0\right) =0$ and
$A_{3}\left( 0\right) =1$.}
\label{fig2}
\end{figure}
%***************************************************************
%***************************************************************
\begin{figure}[!htb]
\centering
\includegraphics[width= 1.0\columnwidth]{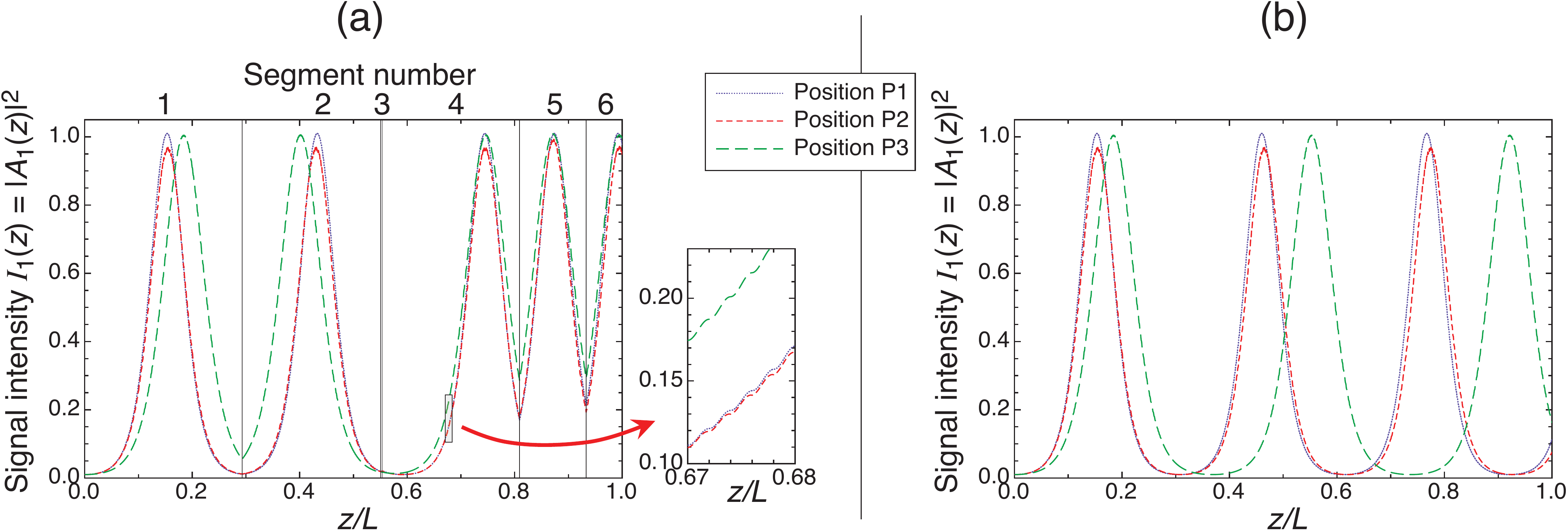}
\caption{(Color online) Spatial evolution of the signal wave intensity for the three points indicated as P1, P2 and P3 in Fig. \ref{fig2} for the cases of a segmented composite crystal (a) and of QPM only (b). The input wave amplitudes are as in Fig. \ref{fig2}. In (a) the segment boundaries are indicated by vertical lines. The zoom in (a) shows the oscillations in the intensity due to each individual QPM periodic domain. Towards the end of the crystal, by the effect of the discontinuities the evolutions get essentially in phase for the three positions in the case (a) but not in the case (b). The parameters are,  position P1: $\widetilde{\Omega}=24/L$, $\Delta k=0$; position P2: $\widetilde{\Omega}=24/L$, $\Delta k=10/L$; position P3: $\widetilde{\Omega}=20/L$, $\Delta k=3/L$. For both graphs the quasi-phase matching period was chosen to be $\Lambda=L/250$. The three-wave mixing equations were integrated using the full version of Eq. (\ref{three-wave mixing}) that takes into account the effect of each periodic domain (see text). }
\label{Fig2bis}
\end{figure}
%*************************************************************

Figure \ref{fig2} illustrates the performance of the composite crystals with
six segments (6a from Table \ref{Table}) compared to standard periodic design.
The amplification values are calculated numerically from Eqs. (\ref{three-wave mixing})
in the cases when $A_{1}\left( 0\right) =0.1$, $A_{2}\left( 0\right) =0$ and
$A_{3}\left( 0\right) =1$ and thus a pump-to-signal intensity ratio $r \equiv I_3(0)/I_1(0)=100$. Figure \ref{fig2} shows clearly that the region
of high signal intensity amplification expands strongly for the segmented composite crystal compared to
standard periodic design. In other words, the composite crystals exhibit
much broader acceptance bandwidths compared to a standard
quasi-phase-matching. The working principle of the composite concept can be recognized directly with the help of Fig. \ref{Fig2bis}, which depicts the evolutions of the signal wave intensity for the conditions associated to the three positions P1, P2 and P3 given in Fig. \ref{fig2}. Clearly each segment boundary gives rise to a "kick" for such evolutions. By choosing the $z$-positions of the frontiers appropriately, one can achieve that the spatial evolutions corresponding to points in the big red area in Fig. \ref{fig2}(a) get very close (nearly "in phase") near the end of the device at distance $L$ with a high final signal wave amplification, as seen in Fig. \ref{Fig2bis}(a). In contrast, for pure QPM without composite segments such a "re-phasing" cannot occur, as can be recognized in \ref{Fig2bis}(b).

The simulations of Fig. \ref{fig2} and Fig. \ref{Fig2bis}
illustrate the general approach for finding the best composite sequence as compared to standard QPM. Next we will prove the concept further by applying these results to specific practical examples and we make the numerics for a real crystal: 5
mol. \% Magnesium Oxide doped Lithium Niobate (MgO:LiNbO$_{3}$). This
ferroelectric nonlinear crystal possesses higher damage threshold compared
to undoped LiNbO$_{3}$, high nonlinear optical coefficient, broad transparency
range and is suitable for domain poling \cite{Nikogosyan}. We compare the
standard quasi phase matching with the composite approach for OPA when the
three interacting beams share the same extraordinary polarisation (Type 0
configuration, all beams polarized parallel to crystal $c$-axis) associated to the largest element of the
nonlinear tensor $d_{333}=\chi^{(2)}= 27$ pm/V.
%***************************************************************
\begin{figure}[!htb]
\centering
\includegraphics[width= 0.7\columnwidth]{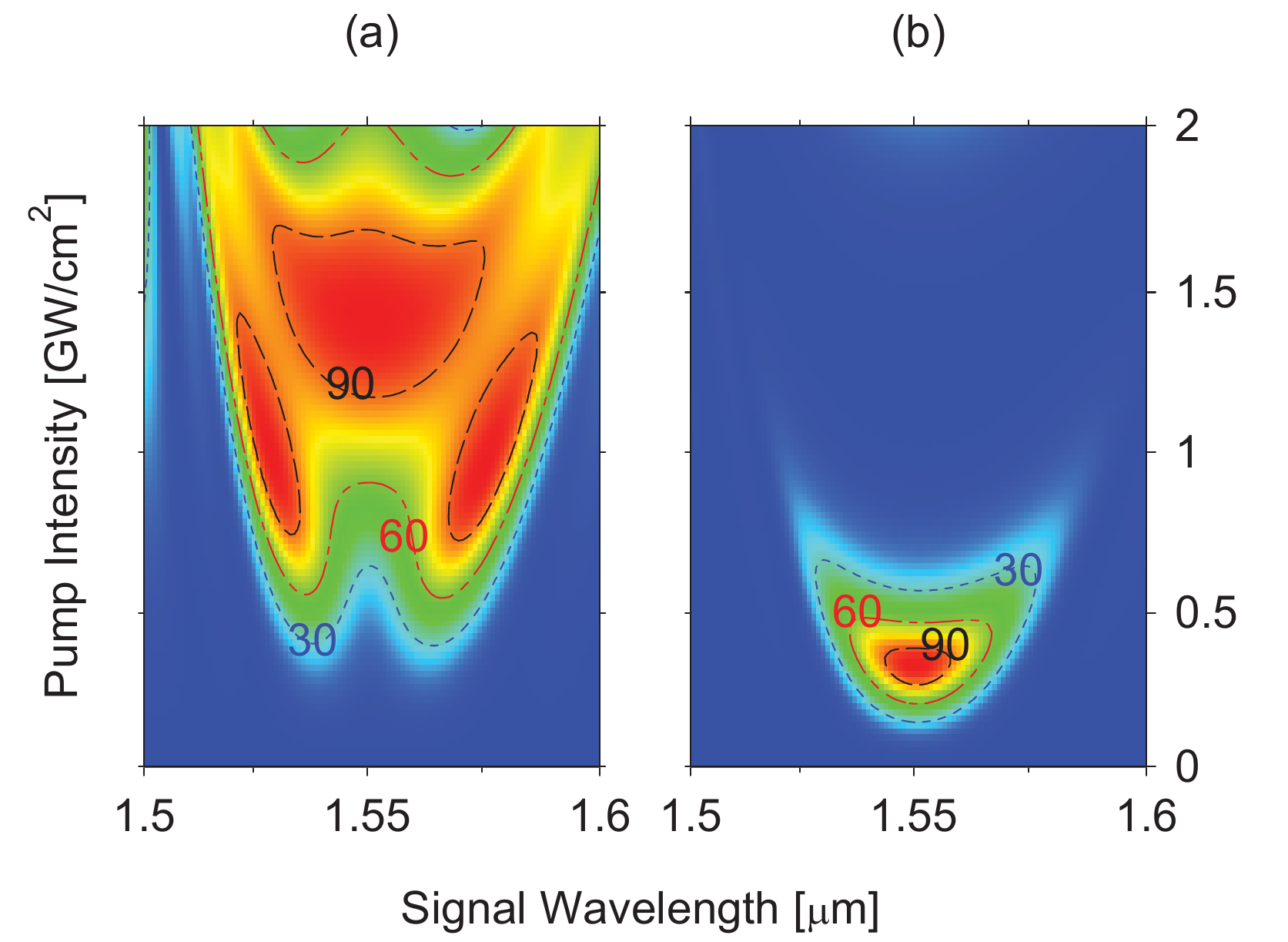}
\caption{(Color online) Color plots of the signal intensity amplification vs. the input pump intensity $I_3$ and the
signal wavelength $\lambda_1$ for the case of MgO:LiNbO$_{3}$ and with initial conditions $A_{1}(0) =0.1$, $A_{2}(0) =0$ and $A_{3}(0) =1$ so that $r=100$. The total crystal length is $L=5$ mm. A pump intensity of 1 GW/cm$^2$ corresponds to a modified coupling coefficient $\Omega =$ 1.32 1/mm at the central wavelength of 1.55 $\mu$m.  (a) Composite crystal with six segments (6b fromTable \protect\ref{Table}). (b) QPM periodic design only. In both cases the poling period is $\Lambda=$29.71 $\mu$m. The red color indicate areas with high signal intensity amplification, while blue color indicate areas with low signal intensity amplification. The three isolines mark intensity amplification level of 30, 60 and 90. }
\label{fig3}
\end{figure}
%***************************************************************
\begin{figure}[!h]
\centering
\includegraphics[width= 0.9\columnwidth]{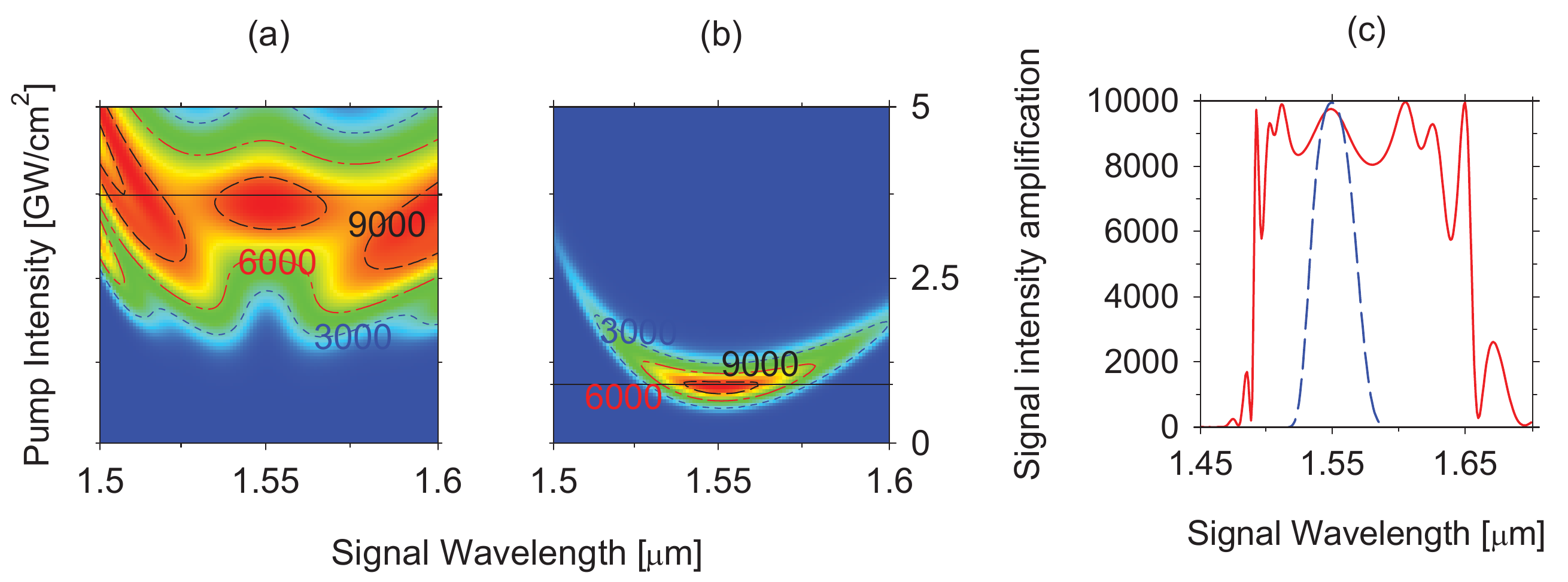}
\caption{(Color online) (a)+(b) Color plots of the signal intensity amplification vs. the input pump intensity $I_3$ and the
signal wavelength $\lambda_1$ for the case of MgO:LiNbO$_{3}$. The initial conditions are $A_{1}(0) =0.01$, $A_{2}(0) =0$ and $A_{3}(0) =1$ so that $r=10000$ and the remaining parameters are as in Fig. \ref{fig3}. (a) Composite crystal with six segments (6c from
Table \protect\ref{Table}). (b) QPM periodic design only. The red (blue) colors indicate areas with high (low) signal intensity amplification. The three isolines mark intensity amplification levels of 3000, 6000 and 9000. The right-hand panel (c) show the signal amplification spectrum for the two cases at the optimum level of pump intensity $I_3$ corresponding to the two horizontal lines in panels (a) and (b), $I_3 = 3.75$ GW/cm$^{2}$ and $I_3 = 0.8$ GW/cm$^{2}$, respectively. The red solid line is for the segmented composite crystal while the blue dashed line for standard QPM only.}
\label{fig4}
\end{figure}
%***************************************************************
%***************************************************************
\begin{figure}[!h]
\centering
\includegraphics[width= 0.7\columnwidth]{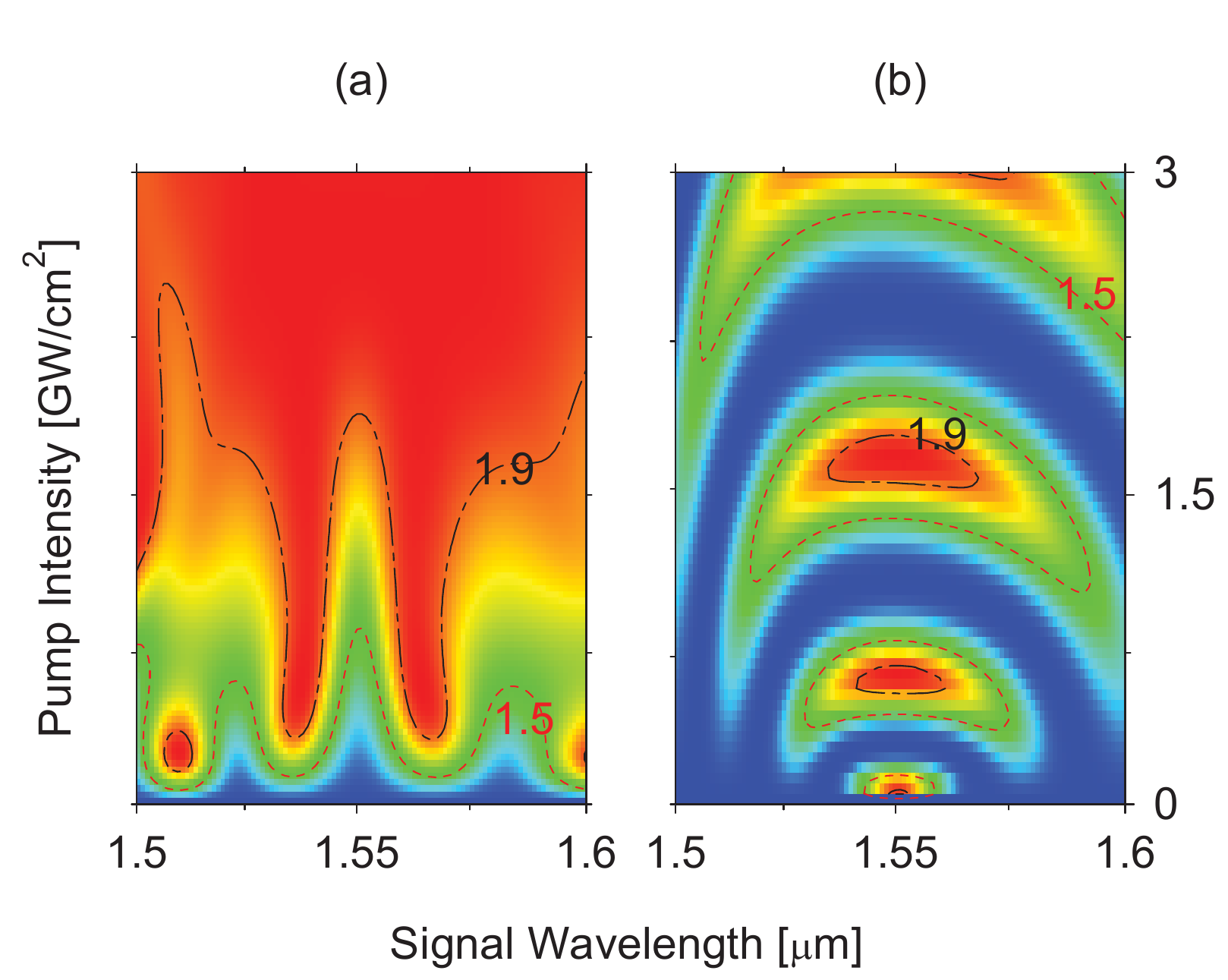}
\caption{(Color online) Color plots of the signal intensity amplification vs. the input pump intensity $I_3$ and the
signal wavelength $\lambda_1$ for the case of MgO:LiNbO$_{3}$. The initial conditions are $A_{1}(0) =1$, $A_{2}(0) =0$ and $A_{3}(0) =1$ so that $r=1$ and the remaining parameters are as in Fig. \ref{fig3}. (a) Composite crystal with eight segments (8 from
Table \protect\ref{Table}). (b) QPM periodic design only.  The red (blue) colors indicate areas with high (low) signal intensity amplification. The two isolines mark intensity amplification levels of 1.5 and 1.9, respectively. }
\label{fig6}
\end{figure}
%***************************************************************

The color plots in Figs. \ref{fig3} and \ref{fig4} compare the signal intensity amplification $a$\
for MgO:LiNbO$_{3}$ for standard QPM ($\Lambda=29.71$ $\mu$m) and for a composite
crystal made of six segments. Figure \ref{fig3} is for intermediate pump-to-signal intensity ratio $r$ while Fig. \ref{fig4} is for large $r$. The nonlinear susceptibility is fixed and the plots are represented for varying input pump intensity (at the fixed wavelength of 1064 nm) and for varying signal wavelength (with center at
$\lambda_1$=1550 nm). Note that here, in order to keep the ratio $r$ constant for each plot, the input signal intensity changes in the same way as the input pump intensity. Note also that the pump intensity ($y$-axis) takes the role of the coupling in Fig. \ref{fig2}. Similarly, since a variation of the signal wavelength with respect to the central one gives rise to a phase mismatch $\Delta k$, here the signal wavelength ($x$-axis) takes the role of $\Delta k$ in Fig. \ref{fig2}. The total crystal length is $L=5$ mm.  The
amplification values are calculated numerically from Eqs. (\ref{three-wave mixing}) in
the cases when $A_{1}(0) =0.1$, $A_{2}(0) =0$ and $%
A_{3}(0) =1$ for Figure \ref{fig3} and $A_{1}(0)=0.01$, $A_{2}(0)=0$ and $%
A_{3}(0)=1$ for Figure \ref{fig4}. Clearly, a greatly enhanced robustness
and frequency bandwidth of the composite OPA compared to the standard QPM OPA can be recognized. The plots in Figure \ref{fig4}(c) compare directly the signal intensity amplification spectrum for the optimal range for composite crystal (red line) and optimal range for standard periodic design (blue dash line). These are slices at the pump intensity values of 3.75 GW/cm$^{2}$ and 0.8 GW/cm$^{2}$ in Fig. \ref{fig4}(a) and Fig. \ref{fig4}(b), respectively.
%***************************************************************

Finally, Fig. \ref{fig6} give the color plots for the signal intensity amplification $a$ as in Fig. \ref{fig3} and Fig. \ref{fig4}, but for the case where the initial signal intensity is as strong as the pump ($r=1$, $A_{1}(0)=1$, $A_{2}(0)=0$ and $%
A_{3}(0)=1$). This is a highly depleted regime and the real advantages of the segmented composite approach over the standard QPM in the depleted pump regime can be recognized by the very large red area in the left panel. In general a laser beam spot has an intensity distribution with strong intensity in the center of the spot and smaller intensity at the wings (Gaussian beam for example). Therefore averaging the amplification value for the big red island of the left panel (a) in Fig. \ref{fig6} will lead to a significantly higher average amplification than averaging the amplification values for the oscillation islands seen for standard QPM in Fig. \ref{fig6}(b).
We can conclude from Figs. \ref{fig3}, \ref{fig4} and \ref{fig6} that the present composite OPA approach works very well in the depleted pump case because in all these cases one has a significant pump depletion and thus a signal amplification approaching the maximum theoretically possible.

%%%%%%%%%%%%%%%%%%%%%%%%%%%%%%%%%%%%%%%%%%%%%%%%%%%%%%%%%%%%%
\section{Summary and Conclusions}

In summary, we have used the similarity between the three wave mixing
equations and the time-dependent Schr\"{o}dinger equation in order to
transfer concepts from quantum physics to nonlinear optics. Specifically, we have suggested to use segmented
composite crystals for optical parametric amplification in analogy with the
composite pulses in NMR and quantum optics. The approach used here is based on
sign-alternating dual-compensating composite pulse sequences similar to those of Shaka and
Pines \cite{Shaka1985,Shaka1987}. These are particularly suited for optical parametric
amplification because besides the standard quasi-phase-matching they require
only additional sign flips of the nonlinear optical susceptibility at specific locations corresponding to the segment frontiers. We have demonstrated numerically that this technique is especially powerful for broadband OPA. The present approach does not require very long crystals and, for the given example of MgO:LiNbO$_3$, is compatible with pump intensities significantly below the damage threshold for ps or sub-ps illumination \cite{Bach17,Meng16}.

\section*{Acknowledgment}
This research was funded by the EU Horizon-2020 ITN project LIMQUET (contract number 765075) and by the Bulgarian Science Fund Grant No. DN 18/14.

\end{document}